# Quantification of sand fraction from seismic attributes using Neuro-Fuzzy approach

Akhilesh K. Verma[1], Soumi Chaki[2], Aurobinda Routray[2], William K. Mohanty[1*],
Mamata Jenamani[3]

[1]Department of Geology and Geophysics, Indian Institute of Technology Kharagpur

[2]Department of Electrical Engineering, Indian Institute of Technology Kharagpur

[3]Department of Industrial and Systems Engineering, Indian Institute of Technology Kharagpur

[*]Email address of corresponding author: wkmohanty@gg.iitkgp.ernet.in

## ABSTRACT

In this paper, we illustrate the modeling of a reservoir property (sand fraction) from seismic attributes namely seismic impedance, seismic amplitude, and instantaneous frequency using Neuro-Fuzzy (NF) approach. Input dataset includes 3D post-stacked seismic attributes and six well logs acquired from a hydrocarbon field located in the western coast of India. Presence of thin sand and shale layers in the basin area makes the modeling of reservoir characteristic a challenging task. Though seismic data is helpful in extrapolation of reservoir properties away from boreholes; yet, it could be challenging to delineate thin sand and shale reservoirs using seismic data due to its limited resolvability. Therefore, it is important to develop state-of-art intelligent methods for calibrating a nonlinear mapping between seismic data and target reservoir variables. Neural networks have shown its potential to model such nonlinear mappings; however, uncertainties associated with the model and datasets are still a concern. Hence, introduction of Fuzzy Logic (FL) is beneficial for handling these uncertainties. More specifically, hybrid variants of Artificial Neural Network (ANN) and fuzzy logic, i.e., NF methods, are capable for the modeling reservoir characteristics by integrating the explicit knowledge representation power of FL with the learning ability of neural networks. In this paper, we opt for ANN and three different categories of Adaptive Neuro-Fuzzy Inference System (ANFIS) based on clustering of the available datasets. A comparative analysis of these three different NF models (i.e., Sugeno-type fuzzy inference systems using a grid partition on the data (Model 1), using subtractive clustering (Model 2), and using Fuzzy c-means (FCM) clustering (Model 3)) and ANN suggests that Model 3 has outperformed its counterparts in terms of performance evaluators on the present dataset. Performances of the selected algorithms are evaluated in terms of correlation coefficients (CC), root mean square error (RMSE), absolute error mean (AEM) and scatter index (SI) between target and predicted sand fraction values. The achieved estimation accuracy may diverge minutely depending on geological characteristics of a particular study area. The documented results in this study demonstrate acceptable resemblance between target and predicted variables, and hence, encourage the application of integrated machine learning approaches such as Neuro-Fuzzy in reservoir characterization domain. Furthermore, visualization of the variation of sand probability in the study area would assist in identifying placement of potential wells for future drilling operations.

1. Introduction

Reservoir characterization of a complex geological system involves the analysis and integration of all available datasets- mainly well log and seismic to finally construct 3D geocellular models of lithological properties. It is important to describe the status of the hydrocarbon reserves from the existing datasets; however, volumetric prediction of the reservoir properties from limited number of well logs and seismic data is viable (Hampson et al., 2001). Literature study associates the growth in reservoir characterization domain with significant advancement in the application of expert systems in the last few decades (Mohaghegh et al., 1996; Fung et al., 1997; Ouenes, 2000; Aminzadeh et al., 2000; Nikravesh, 2001; Nikravesh et al., 2003; Lim, 2003; Nikravesh, 2004; Hou et al., 2008; Kadkhodaie-Ilkhchi et al., 2009; Haixiang, et al., 2011; Kaydani and Mohebbi, 2013; Al-Dousari and Garrouch, 2013). The objective of this type of studies is to first attune a functional relationship between reservoir properties and predictor variables at available well locations, and then apply the derived relationship to estimate the target variable over the study area from seismic attributes. Inclusion of 3D seismic attributes in the modeling empowers machine learning algorithms to estimate target properties away from the wells (e.g., Hilterman, 1999; Hampson et al., 2001; Hou et al., 2008; Kadkhodaie-Ilkhchi et al., 2009). The reliability of the prediction model needs to be ensured by means of testing the prediction algorithms using unseen validation set at well control points.

Geophysical well log data provides a high vertical resolution of different physical properties of the lithological units. However, in-situ measurements of these properties are difficult to perform due to limited spatial accessibility and hence, most of the reservoir properties are derived from indirect measurements. The analyzed core data are normally used to guide the derivation process by comparing it with the recorded logs in the vicinity of the existing wells. On the other hand, seismic data provides higher horizontal resolution conceding the vertical resolution level to some extent. Main uses of seismic data are in the delineation of subsurface structures, and hence reservoir bodies; however less helpful in the essential task of quantifying spatial distribution of reservoir properties. Therefore, quantification of inherent relationship between seismic and log properties at the well locations provides a direction to integrate these data to get 3D model of reservoir variables. This can be achieved through systematic combined study of well log and seismic data for 3D reservoir modeling. Moreover, delineation of thin reservoir bodied creates several challenges such as availability of high resolution log data at few locations and low resolvability of seismic data. However, as wells cannot drilled everywhere due to much cost and accessibility, the only alternative is to use seismic data as predictor to model reservoir properties away from well and between the wells to get 3D reservoir model. Following paragraph enlists the literatures available in this research field.

In last three decades, several researchers have used soft computing and Artificial Intelligence (AI) techniques for reservoir characterization (RC). Literature study reveals that machine learning algorithms exhibit superior performance in terms of high accuracy and low error value compared to conventional statistical methods (Nikravesh and Aminzadeh, 2001). Some of the important articles with special emphasis on RC using 3D seismic data include Boadu, (1997); Nikravesh, (1998a,b); Nikravesh et al., (1998); Chawathe et al., (1997); Yoshioka et al., (1996); Schuelke et al., (1997); Monson and Pita, (1997); Aminzadeh and Chatterjee, (1985); Finol and Jing (2002); Hou et al., (2008); Kadkhodaie-Ilkhchi et al., (2009); Al-Dousari and Garrouch (2013); Na'imi, et al., (2014); Iturrarán-Viveros and Parra (2014). Other important works on application of Artificial Neural Network (ANN), Fuzzy Logic (FL) and Neuro-Fuzzy (NF) methods for RC from well log data and core data are progressively reported by many authors (Baldwin et al., 1989, 1990; Pezeshk et al., 1996; Rogers et al., 1992; Wong et al., 1995a, b; Nikravesh et al., 1996; Nikravesh and Aminzadeh, 1997; Klimentos and McCann, 1990; Aminzadeh et al., 1994; Huang and Williamson, 1994; Griffiths, 1987; Adams et al., 1999; Bhatt and Helle, 2002; Lim, 2005; Haixiang, et al., 2011; Nooruddin et al., 2014). Many research papers have been documented on reservoir characterization using statistical approaches (Doyen, 1988; Majdi et al., 2010) and its integration with neural network (Aminzadeh et al., 2000; Al-Dousari and Garrouch, 2013; Na'imi, et al., 2014; Sauvageau et al., 2014). Detailed discussion on neural network and its application is well documented in a review article by Poulton, (2002). The concept of fuzzy logic is useful in handling uncertainty associated with nonlinear mapping (Ouenes, 2000; Nikravesh and Aminzadeh, 2001; Kadkhodaie-Ilkhchi et al., 2009). Several studies suggest that hybrid approaches such as Neuro-Fuzzy methods, Neuro-Genetic methods, Neuro-Fuzzy-Genetic methods would be effectively applied for the accurate modeling of nonlinear variables (Nikravesh and Aminzadeh, 2001; Dorrington and Link, 2004; Saemi et al., 2007; Kadkhodaie-Ilkhchi et al., 2009; Kaydani and Mohebbi, 2013; Ahmadi, et al., 2013; Nooruddin, et al., 2014). However, customization of non-linear methods may be required to obtain optimum performance depending on nature of the working dataset. Kadkhodaie-Ilkhchi et al., (2009) describes a committee fuzzy inference system for prediction of petrophysical properties. Three inference systems namely Mamdani, Larsen and Sugeno are used. Most of the studies are focused to either porosity or water saturation or both. Moreover, maximum number of papers is concentrated on the modeling of reservoir properties using well log data. However, a few have discussed on effectiveness of seismic data along with logs for generating reservoir models (Hou et al., 2008; Kadkhodaie-Ilkhchi et al., 2009; Al-Dousari and Garrouch, 2013).

In the present study, the reservoir constitutes very thin layers of sand and shale which are difficult to discriminate from impedance log (Fig. 1). Analysis and interpretation of such reservoirs having thin layers is possible by accurate modeling of reservoir properties such as sand fraction or shale fraction (content) from integrated study of well log and seismic attributes. Moreover, a systematic study on the integration of different dataset and application of hybrid machine learning algorithms is rare in existing literature. Thus, the present study concentrates on the assessment of potential of the NF approaches and its comparison with ANN in the modeling of reservoir property. In this study, a hybrid approach of neural network and fuzzy logic is applied where three different inference systems are used to model sand fraction from seismic attributes. Quantification of sand/shale fraction for a reservoir system provides crucial information about sand/shale content, and hence potential reservoir.

The contribution of this study are listed as: 1) assimilation of 3D seismic data along with well logs for modeling of reservoir property; 2) assessment of thin sand layers by proposing a hybrid technique of neural network and fuzzy logic to model sand fraction from seismic attributes; in which three different categories for Adaptive Neuro-Fuzzy Inference System (ANFIS) are used; 3) comparison has been made of the results obtained from three different NF models and ANN, where performance metrics are CC (Correlation coefficient), RMSE (Root mean square error), AEM (Absolute Error Mean) and SI (Scatter Index).

## 2. Description of data

In this study, the well log and seismic data are acquired from a hydrocarbon producing field located in the western onshore of India. The borehole dataset contains geophysical logs such as gamma ray, resistivity, density and other derived logs, e.g., sand fraction value, porosity, water saturation, etc. On the other hand, the seismic dataset includes different attributes, i.e., seismic amplitude, seismic impedance, instantaneous frequency, seismic envelope, seismic sweetness etc.; these attributes are mainly derived from seismic amplitude. Present study aims to model a petrophysical property from seismic attributes. In this context, choice of relevant predictor attributes from a set of attributes is a crucial job to achieve good results within a less execution time. Literature review offers multiple data mining algorithms to determine relevant features from available "candidate features" to model or classify a target variable. Some of them are mutual information (MI) (Oveisi et al. 2012), normalized mutual information (NMI) (Estévez et al. 2009), gamma test (Jones 2004; Noori et al. 2011; Iturrarán-Viveros, 2012), profile method (Gevrey et al., 2003), Relief algorithm (Kira and Rendell, 1992; Chaki et al. 2014a), principal component analysis (PCA), forward selection (FS) techniques (Gevrey et al. 2003), sequential forward feature selection (Safarzadegan et al., 2012). In the present study, we have opted for a sequential forward feature selection technique to obtain the most significant input attributes from above mentioned five seismic attributes (Gilan et al., 2012). Here, predictor attributes are added sequentially to an "empty candidate set", where at each stage the variable sets with the best performance is used in the next stage. The stopping criterion of addition of variables at each consecutive stage is decrease in performance indicator with the addition of further variables in the prediction model. We have selected correlation coefficient (CC) to decide the performance of the prediction model with different sets of input variables. Fig. 2 represents the predictor selection results for the prediction of sand fraction for Well A in terms of CC using an ANN having single hidden layer. It reveals that the best performance is achieved when seismic impedance, seismic amplitude, and instantaneous frequency are used in the model as input attributes to predict sand fraction. These three predictor attributes have yield good performance in prediction of sand fraction in previous studies also (Chaki et al. 2014b). Therefore, three seismic attributes (i.e., seismic impedance, amplitude, and instantaneous frequency) are used to build NF models to predict sand fraction. Here, an integrated dataset of seismic attributes and sand fraction at available six well locations named as A, B, C, D, E and F is used. With such dataset, we started analyzing impedance variations of sand and shale units, and found that they overlap to each other. Fig. 1 reveals that the histogram plots of impedance corresponding to sand and shale units overlap to each other. Therefore, the identification of thin reservoir layers of sand and shale is difficult from impedance information only. Cross plots of the target petrophysical property (sand fraction) with the three seismic attributes (seismic impedance, seismic amplitude and instantaneous frequency) are demonstrated in Fig. 3. The low correlation values between the target and predictor variables are evident in these figures which anticipate the challenge in the modeling of reservoir properties from seismic attributes. In addition to this, the cross plots between seismic attributes and sand fraction do not (Fig. 3) reveal any cluster, and hence do not help in discriminating different lithologies.

3. Methodology

Concepts of neural networks and fuzzy logic have been used to solve nonlinear problems in difference fields of science and technology (Nikravesh et al., 2001; Li and Tong, 2003; Chairez, 2013; Fung et al., 1997). Particularly, ANNs have gained immense popularity for clustering, nonlinear mapping and classification of dataset (Fung et al., 1997; Nikravesh et al., 2001). The building blocks of ANN consist of input, hidden and output layers of neurons (Haykin, 1999). A single neuron is characterized by input weights, a threshold and a linear or nonlinear activation function. A neuron in one layer is connected to other layers by weights in a feed-forward manner. Thus, the output of a particular neuron is a non-linear function of the weighted sum of the inputs to the neuron plus a bias. The major issues to train the neural networks are: initialization of synaptic weights, network configuration and generalization, over-fitting, under-fitting, independent validation, and speed of convergence.

On the other hand, the concept of fuzzy logic is applied in different research fields for its ability to handle uncertainties associated with the datasets and models (Zadeh, 1965; Chairez, 2013). The idea of fuzzy logic or fuzzy sets theory, introduced by Zadeh, in 1965, defines the partial belongingness of an element to a set rather than it is completely belonging to that set or not, as in classical (crisp) set theory. Fuzzy sets allow assigning a degree of membership to the elements in that set. In other words, the transformation from "belong to a set" to "not belong to a set" is gradual in fuzzy sets. Thus, a fuzzy set *A* in *X* (a universe of discourse of elements x) can be defined as

$$A = \{(x, \mu_A(x)) | x \in X\} \quad (1)$$

A fuzzy set consists of two things: a) Element, and b) Membership value, and degree of membership in set *A* is defined by its membership function as given below

$$\mu_A : X \rightarrow [0,1] \quad (2)$$

Membership function maps each element of *X* to a membership value between 0 and 1; i.e., it is a representation of belongingness of each element of *X* to set *A*. Moreover, detailed representation of fuzzy logic (fuzzy sets, fuzzy membership function, fuzzy if-then rules, and operation in fuzzy sets, etc.,) can be found in well referred literatures like Zadeh (1965), Kosko (1992), and Jang et al., (1997).

In the exploration geosciences, the datasets are acquired from a complex and heterogeneous system, and hence, they exhibit nonlinear characteristics. Concisely, modeling of reservoir variables from available data requires some knowledge based inputs to deal with the uncertainty. Knowledge to the model can be provided through linguistic representation of the variables. Number of fuzzy partitions is decided based on the classification of the data set. In the present study, sand fraction value is classified into three classes, i.e., low, medium, and high. Moreover, the stratigraphic changes of different depositional layers follow the smooth variation in the physical properties of different lithological units. There are several membership function types such as Gaussian, generalized bell-shape, trapezoidal, sigmoid, triangular, etc., that can used in the fuzzy inference system. However, in the present case study, bell-shape membership function is used empirically, which yields good performance. A typical example of the representation of fuzzy membership function (bell-shape) for sand fraction is presented in Fig. 4, where x-axis represents the values of sand fraction and corresponding membership grades are denoted on y-axis. Here, a crisp sand fraction value 0.12 has membership grade values 0.136 and 0.7147 correspond to predefined medium and low membership functions respectively. The sand fraction value 0.12 has no belongingness to high membership function as shown in Fig. 4.

In the context of integrated approach, Adaptive Neuro-Fuzzy Inference System (ANFIS) is a hybridization of neural network (Haykin, 1999) and fuzzy logic (Zadeh, 1965; Jang, 1993). Generally, the nucleus of a fuzzy system is the Fuzzy Inference System (FIS). This inference mechanism is carried out by a set of rules operating upon variables which are fuzzified through various membership functions. The number and type of fuzzy rules, membership functions and the method of inference are the basic building blocks of the FIS system. Fuzzy if-then rules or fuzzy conditional statements are expressions of the form IF A THEN B, where A and B are labels of fuzzy sets characterized by appropriate membership functions. Due to their concise form, fuzzy if-then rules are often employed to capture the imprecise modes of reasoning that play an essential role in the human ability to make decisions in an environment of uncertainty and imprecision. Basically a fuzzy inference system is composed of five functional blocks: 1) a rule base containing a number of fuzzy if-then rules; 2) a database which defines the membership functions of the fuzzy sets used in fuzzy rules; 3) a decision-making unit which performs the inference operations on the rules; 4) a fuzzification interface which transforms the crisp inputs into degrees of match with linguistic values; and 5) a defuzzification interface which transform the fuzzy results of the inference into crisp output.

ANFIS has been widely used in many fields of science and technology (Li and Tong, 2003; Chairez, 2013; Chaki et al., 2013). In this study, ANFIS is applied to model a lithological property (sand fraction) from seismic attributes. Outline of the modeling approach using ANFIS is presented in Algorithm 1. Detailed descriptions of the steps involved in this algorithm are demonstrated in subsequent sections.

| **Algorithm 1**: Modeling using Neuro-Fuzzy Approach |
| --- |
| **Input**: Predictor and target attributes in depth or time domain |
| **Output**: A heuristic mapping of target property and input predictors |
| 1: **for** $i$ = 1 to $n$ ($n$: number of data point) |
| 2: Division of input-target data pairs ($d_i$) into two parts, i.e., training and testing |
| 3: Data normalization |
| 4: **end for** |
| 5: **for all** $d_i$ pairs **do** |
| 6: Fuzzification of inputs by assigning membership function (MF); e.g. generalized bell-shaped MF which can be expressed as: $$f(x;a,b,c) = \frac{1}{1+\left|\frac{x-c}{a}\right|^{2b}}$$ where $c$ represents the center of the corresponding membership function, $a$ represents half of the width; $a$ and $b$ together determine the slopes at the crossover points. This function is continuous and infinitely differentiable. |
| 7: Initialization of Fuzzy Inference System (FIS) |
| 8: Creates a fuzzy decision tree to classify the data into one of $2^m$ (or $p^m$) linear regression models to minimize the sum of squared errors; where $p$ is the number of fuzzy partitions of each variable and $m$ is the number of input variables |
| 9: Assign number of epochs |
| 10: **for** 1 to $j$ ($j \leq$ number epoch) |
| 11: Fuzzy If-then-rules and output normalization |
| 12: Modification of FIS |
| 13: Defuzzification of output |
| 14: Calculate root mean square error |
| 15: This is done for both training and testing data |
| 16: Save the network structure and parameters for minimum error |
| 17: **end for**; **end for** |
| 18: Apply the obtained inputs-target relationship to the inputs off from the training-testing data |

*3.1 Takagi-Sugeno-Kang fuzzy system and its implication*

The computational process of ANFIS is based on the Sugeno fuzzy model proposed by Takagi, Sugeno and Kang (Jang, 1993; Takagi and Sugeno, 1985) aiming to construct if-then fuzzy rules from an input-output dataset. Fig. 5 (a) illustrates a fuzzy reasoning for two inputs and one output, and its equivalent ANFIS architecture (Takagi and Sugeno's ANFIS) is shown in Fig. 5(b). Processing steps are shown in five dotted boxes (Input, Fuzzification, Inference process, Defuzzification and output). It can be seen from Fig. 5(b) that ANFIS consists five layers whose functions are briefly discussed below (Jang, 1993):

**Layer 1:** Every node in this layer generates membership grades of a linguistic label (small, large etc.) and the parameters in this layer are referred to as the premise parameters. It can be represented as

$$O_i^1 = \mu_A(x) \tag{3}$$

Where $O_i^1$ is the membership function of $A_i$ and it specifies the degree to which the given $x$ satisfies the quantifier $A_i$. Most prefer membership function is bell-shaped with maximum equal to 1 and minimum equal to 0. The function is continuous and infinitely differentiable (See the step 6 of above mentioned Algorithm 1). Therefore, generalized bell shaped membership function is selected as learning may go from +∞ to –∞ (Chaki et al., 2013).

**Layer 2:** Every node in this layer calculates the firing strength of a rule via multiplication operation as shown below.

$$w_i = \mu_A(x) \times \mu_B(y); \; i = 1, 2 \tag{4}$$

**Layer 3:** Every $i^{th}$ node in this layer calculates the ratio of the $i^{th}$ rule's firing strength to the total of all firing strength. For $i^{th}$ node, it can be represented as

$$\overline{w}_i = \frac{w_i}{w_1 + w_2}, i = 1, 2 \tag{5}$$

**Layer 4:** Every $i^{th}$ node in this layer computes the contribution of the $i^{th}$ rule to the overall output.

$$o_i^4 = \overline{w}_i f_i = \overline{w}_i (p_i x + q_i y + r_i) \tag{6}$$

where $\{p_i, q_i, r_i\}$ is the consequent parameters.

**Layer 5:** The single node in this layer computes the overall output as the summation of contribution from each rule.

$$o_i^5 = overall\,output = \sum_i \bar{w}_i f_i = \frac{\sum_i w_i f_i}{\sum_i w_i} \tag{7}$$

In this way, an adaptive network equivalent to a Takagi and Sugeno's fuzzy inference system is constructed. From the ANFIS architecture, it is observed that given the values of premise parameters, the overall output can be expressed as linear combinations of the consequent parameters. More precisely, the output $f$ in Fig. 5(a) and Fig. 5(b) can be written as

$$\begin{aligned} f &= \frac{w_1 f_1}{w_1 + w_2} + \frac{w_2 f_2}{w_1 + w_2} \\ &= \bar{w}_1 f_1 + \bar{w}_2 f_2 \\ &= (\bar{w}_1 x) p_1 + (\bar{w}_1 y) q_1 + (\bar{w}_1) r_1 + (\bar{w}_2 x) p_2 + (\bar{w}_2 y) q_2 + (\bar{w}_2) r_2 \end{aligned} \tag{8}$$

This represents a linear relation between consequent parameters ($p_1$, $q_1$, $r_1$, $p_2$, $q_2$ and $r_2$).

*3.2 Clustering of data*

Literature study reveals several clustering techniques such as K-mean clustering, fuzzy C-mean clustering, subtractive clustering, probability density function, Neuro-Fuzzy based clustering, etc. These clustering algorithms are used extensively not only for organization and categorization of dataset, but also for data compression and model construction. This section discusses some of the mostly used clustering techniques in conjunction with radial basis functional networks and fuzzy modeling: c-means (or K-means) clustering, fuzzy K-means clustering and subtractive clustering.

The K-means algorithm (MacQueen, 1967) partitions a collection of n vectors $X_j$, $j = 1, 2...n$, into $c$ groups $G_i$, $i = 1... c$, and finds a cluster center in each group such that a cost function of dissimilarity (distance) measure is minimized.

On the other hand, Fuzzy c-means clustering (FCM) is a clustering algorithm in which each data point belongs to a cluster to a degree specified by a membership grade (Bezdek, 1981). In this technique, a collection of *n* vector $X_j$, $j = 1, ..., n$ is partitioned into *c* fuzzy clusters. Then a cluster center is determined in each cluster by minimizing the cost function. In FCM, the fuzzy partition is done in such a manner that a given data point can belong to several clusters with the degree of belongingness specified by membership grades between 0 and 1. In such cases, the membership matrix *U* is allowed to have elements with values between 0 and 1. However, imposing normalization stipulates that the summation of degrees of belongingness for a dataset always be equal to unity:

$$\sum_{i=1}^{c} u_{ij} = 1, \forall j = 1, 2, ..., n \tag{9}$$

The cost function for FCM is given below.

$$J(U, c_1, ..., c_c) = \sum_{i=1}^{c} J_i = \sum_{i=1}^{c} \sum_{j}^{n} u_{ij}^m d_{ij}^2 \tag{10}$$

where $U_{ij}$ is between 0 and 1; $C_i$ is the cluster center of fuzzy cluster *i*; $d_{ij} = \|C_i - X_j\|$ is the Euclidean distance between $i^{th}$ cluster center and $j^{th}$ data point.

The cluster centers can also be first initialized and then updated iteratively. The performance depends on the selection of initial cluster centers, thereby allowing the user either to use another fast algorithm to determine the initial cluster centers or to execute FCM several times, each starting with a different set of initial cluster centers. FCM is widely used in the creation of Sugeno type fuzzy inference systems.

In most of the clustering algorithms, it is typically required the user to pre-specify the number of cluster centers, such as in the Fuzzy c-means algorithm. Specifically, in fuzzy modeling, the subtractive clustering (Chiu, 1994) is a widely used technique for determining the number of clusters and cluster centers to create the fuzzy inference system (FIS) (Jang, 1993; Chiu, 1994; Jarrah and Halawani, 2001). The number of clusters determines the number of rules and membership functions in the generated FIS. Subtractive clustering follows the mountain method (Yager and Filev, 1994) for estimating the number and initial location of cluster centers.

It considers each data points as a potential candidate for cluster centers. The procedure of determining the cluster centers using the subtractive clustering is given as (Chiu, 1994):

Suppose, a collection of *n* vectors $x_1$, $x_2$, $x_3$..., $x_n$ represents the data points and each data point is considered as a potential cluster center. The potential of data point $x_i$ is determined as:

$$P_i = \sum_{j=1}^{n} e^{-\frac{4\|x_i - x_j\|^2}{r_a^2}} \tag{11}$$

Where the symbol $\| \cdot \|$ represents the Euclidean distance, and $r_a$ is a positive constant. Thus, the measure of the potential for a data point is a function of its distances to all other data points. The constant $r_a$ is effectively the radius defining a neighborhood; data points outside this radius have little influence on the potential. Thus, the radius of neighborhood plays a crucial role in determination of the number of clusters. After the potential of every data point has been computed, the data point with the highest potential is selected as the first cluster center. Then, an amount of potential is subtracted from each data point as a function of its distance from the first cluster center. In this way, next cluster center is selected based on the potential of data points. This process continues until a sufficient number of clusters are obtained.

In the present study, subtractive clustering (Chiu, 1994) is used to determine the cluster centers for fuzzy modeling. The performance of fuzzy models is dependent on the radius of the cluster, and hence cluster center. The number of clusters is strongly related to the number of fuzzy if-then rules. In this type of clustering approaches, cluster radius, which ranges from 0 to 1, plays a crucial role. Smaller cluster radius leads to relatively more number of clusters compare to the clusters with large radius (Chiu, 1994; Jarrah and Halawani, 2001). In the present case study, the clustering radius is taken as 0.2, which is selected empirically.

Present work represents a study of NF modeling using adaptive fuzzy inference system based on three types of clustering, i.e., partition of data, subtractive clustering, FCM clustering. Subtractive clustering algorithm (Chiu, 1994) is used to determine the cluster centers in last two methods (i.e., NF modeling based on subtractive clustering and FCM clustering). A comparative study of NF methods and ANN is also carried out to observe the advantage of integrated concept of neural network and fuzzy logic.

## 4. Modeling of sand fraction from seismic attributes

In this paper, modeling of sand fraction is carried out from three seismic attributes (seismic impedance, seismic amplitude and instantaneous frequency) using Neuro-Fuzzy approach with three different inference systems. These fuzzy inference system structures are based on gridding, subtractive clustering, and FCM clustering of the dataset, respectively. This is carried out by extracting a set of rules from the available data patterns.

Application of these nonlinear mapping techniques to the real data requires several pre-processing, modeling, and post-processing steps. These steps include data integration, re-sampling, normalization, learning and validation, followed by post-processing.

*4.1 Integration of seismic and well log data (re-sampling)*

The integration of data from different sources with heuristic knowledge from human experts is the critical step of reservoir characterization (Nikravesh, 2004). The task is to combine the seismic attributes at each available well location with the lithological properties. Therefore, first, we extract the seismic values at six well locations from seismic data cube. Seismic attributes are presented in time domain whereas well log data are recoded along the depth. Depth to time conversion of well logs has been carried out using suitable velocity profile resulting from well-to-seismic tie. After converting the well logs from depth to time domain, we have found out that the sampling rate of these two types of signals are different. The sampling interval of band limited seismic signals is two milliseconds, whereas the well logs are sampled at an interval of ~0.15 milliseconds. Since, the sampling interval of both the files are different, we apply Nyquist–Shannon sampling theorem (Shannon, 1949), which states that a band limited signal can be completely reconstructed from the samples, to reconstruct seismic attributes at each time instant corresponding to the well logs using cubic spline interpolation (de Boor, 1978; Neill, 2002).

*4.2 Data normalization*

Data normalization plays a crucial role for tuning the performance of the modeling algorithms. During the experimentation with the current dataset, we propose different normalization schemes for predictor and target variables. The predictor variables are normalized using the Z-score normalization. The values of attribute *X* are normalized using the mean and standard deviation of the *X*. The normalized value is obtained following the equation:

$$normalized\_val = \frac{val - \mu_X}{\sigma_X} \qquad (12)$$

where $\mu_X$ and $\sigma_X$ represent the mean and standard deviation of the attribute *X*.

Then, the target variable is normalized using min-max normalization that performs a linear transformation on original data values. The relationships among the original data values are preserved in this normalization. For an instance, $\min_X$ and $\max_X$ are minimum and maximum values of attribute *X*. This data interval $[\min_X, \max_X]$ is to be mapped into a new interval $[new\_\min_X, new\_\max_X]$. Therefore, every value from original data interval is normalized using the following function.

Consider *f* is a continuous function $f : [\min_x, \max_x] \rightarrow [new\_\min_x, new\_\max_x]$ defined as:

$$f(x) = \frac{x - \min_X}{\max_X - \min_X}(new\_\max_X - new\_\min_X) + new\_\min_X \qquad (13)$$

where *x* is a variable, and f(x) is normalized variable.

*4.3 Data division*

In most of the nonlinear statistical modeling approaches, it is common to divide dataset into training and testing patterns for learning and validation respectively. In the present study, datasets from six wells are used for modeling of reservoir property. Here, predictors are seismic attributes and target data is sand fraction. Combined dataset from six well are divided into two parts, i.e. training and testing; 70% data has been used for training and rest 30% are used for testing. Division of the data is done using random sample selection (Särndal, et al., 2003). Scatter plots depicting low correlation between predictors (seismic attributes) and target log (sand fraction) are presented in the earlier section.

*4.4 Learning and validation*

Learning from the available training samples is governed by training algorithms where relationship between inputs and target is established heuristically. We prepare the master dataset combining seismic attributes and limited number of well logs (six wells in this case). Then, the master dataset is divided into two parts for training (70%) and testing (30%) purpose after randomizing the sample patterns. For each model, initialization of parameters associated with the respective predictor model is carried out heuristically. Then, the initial values are modified systematically keeping the improvement of training and testing performances in view. Training and testing processes have been carried out using ANN and three different variants of NF modeling, i.e. Adaptive Neuro-Fuzzy Inference System based on 1) data partition, 2) subtractive clustering of data, and 3) FCM clustering. Performance of each fuzzy inference structure model and ANN model is discussed in the following section. Furthermore, trained parameters of each case are saved, and further used to populate the sand fraction value across the study area from seismic attributes. Visualizations of predicted sand fraction across specific In-lines are demonstrated.

*4.5 Post processing and visualization*

Predicted sand fraction values throughout the study area reveal scattered variation. It may be due to much variation or discontinuity in the sand reservoir distribution. Therefore, smoothing of blindly predicted sand fraction is required. In reality, the petrophysical properties across the depth cannot change abruptly from a very high value to a lower one and vice versa. The transition must be as smooth as possible. To incorporate this rationale, we filter the predicted values through a 2-D median filter (Huang et al., 1979). We treat each of the inline as an image and run 2-D median filtering algorithm over it. A 3x5 window that slides over the image from left to right is selected. The window corresponding to each sample point considers the other points taking intersection of two nearest cross lines and four depth wise slices. Visualization of post processed data is presented in the form of slices.

*4.6 Performance metrics*

The performance of prediction methods can be assessed using many state-of-art performance metrics such as time complexity, space complexity, and accuracy of the model. These metrics used to evaluate and compare between prediction methods. In the present study, the main concern is accuracy rather than time and space because training stage is performed once. The performance of the trained networks are evaluated using four parameters – correlation coefficient (CC) and root mean square error (RMSE), absolute error mean (AEM), and the scatter index (SI). The scatter index represents the ratio of RMSE to mean of in situ observations (Sharma and Ali, 2013). These statistical characteristics are defined as

Correlation coefficient (CC): $CC = \sum_{i=1}^{N}(X_i - \overline{X_i})(Y_i - \overline{Y_i}) / \sqrt{\sum_{i=1}^{N} X_i - \overline{X_i})^2 (Y_i - \overline{Y_i})^2}$

Root mean square error: $RMSE = \sqrt{\sum_{i=1}^{N}(X_i - Y_i)^2 / N}$

Absolute error mean: $AEM = \frac{1}{N}\sum_{i=1}^{N}|X_i - Y_i| = \frac{1}{N}\sum_{i=1}^{N}|e_i|$

Scatter index: $SI = RMSE / \overline{Y}$

Where, $X_i$ and $Y_i$ ( i=1, 2, 3,···,N) represent modeled and observed values, respectively, $\overline{X}$ and $\overline{Y}$ are their corresponding average values. Total number of data points are $N$. $e_i$ denotes absolute error.

## 5. Results and discussion

This section represents the analysis of the results obtained from training and testing of the NF modeling based on 1) data partition, 2) subtractive clustering of data, and 3) FCM clustering and ANN model. Figs. 6–8 represent the performance of ANFIS based on data partition corresponding to wells B, C and E, respectively. Similar results have been obtained for other wells also. In each case, (a) the superimposed plot of actual (target log) and predicted sand fraction, and (b) the cross plot of actual and predicted values are presented. Similarly, Figs. 9–11 are corresponding to the performance of ANFIS modeling based on subtractive clustering for the same three wells, and finally, Figs. 12–14 demonstrate the results of NF modeling based on FCM clustering. Performances of the selected algorithms are evaluated in terms of correlation coefficients (CC), root mean square error (RMSE), absolute error mean (AEM) and scatter index (SI) between target and predicted sand fraction values. The obtained CCs for six wells are demonstrated in Table I for three cases of Neuro-Fuzzy model (i.e., adaptive NF models based on data partition (Model 1), subtractive clustering (Model 2), and FCM clustering (Model 3)) and ANN. The accuracy of different models is compared using bar plots as represented in Figs. 15, 16, 17 and 18 for CC, RMSE, AEM and SI, respectively. Good matching of target and predicted sand fraction, which is evident from correlation coefficient and other performance evaluators' values, represent the generalization and potential of the proposed technique. Thus, a functional relationship between sand fraction and predictors (seismic attributes) is calibrated, and then, the tuned model parameters are used to estimate the sand fraction across the area from seismic attributes. Close analysis of the results obtained using Neuro-Fuzzy models and ANN model suggest that among all the four models, model 3 performed better for the present dataset.

As the performances of the algorithms are satisfactory in terms of high value of correlation coefficient and low error values, therefore, volumetric prediction of the sand fraction is carried out from seismic attributes over the study area using the trained parameters. Further, the predicted sand fraction volume has been smoothen by applying median filtering to get improved visualization of sand fraction across the study area. Figs. 19–21 represent visualization of volumetric prediction after smoothing at a particular In-line corresponds to well C. The target sand fraction plot of well C is superimposed on the volumetric slice. The predicted sand fraction at well C mostly follows the target sand fraction log presented in black line over the slice. The color code represents the variation of sand fraction across the volume from low to high range. Preliminary analysis of well log data reveals that the reservoir in the study area consist of thin layers of sand and shale, which can also be seen in the 3D model of sand fraction developed in the present study. Modeling of the distribution of sand or shale fraction in reservoir area is a significant contribution for further study toward the identification of hydrocarbon reserves.

## 6. Conclusions

This paper presents the capability of three existing Neuro-Fuzzy (NF) techniques and their comparison with ANN in the modeling of sand fraction (an important reservoir property in petroleum exploration) from seismic attributes using real exploration data from western onshore of India. The reservoir constitutes thin layers of sand and shale, therefore, the issue of modeling sand fraction is addressed. Moreover, 3D geo-cellular model of target property is created from three seismic attributes over the study area. Analysis from performance evaluators of the results obtained using NF models and ANN model suggest that among all the four models, Model 3 performed better for the present dataset. Moreover, comparative study of three categories of NF models indicates the generalization capabilities of the used methods; hence, it recommends further use of integrated approaches in case of other nonlinear mappings involving different parameters. However, performance of the individual technique may vary depending on the inherent characteristics of a different dataset.